\documentclass[showpacs,twocolumn,pra]{revtex4}%
\usepackage{amsfonts}
\usepackage{amsmath}
\usepackage{amssymb}
\usepackage{graphicx}%

\begin{document}

\title{Rashba spin-orbit coupled atomic
Fermi gases}

\author{Lei Jiang$^{1}$, Xia-Ji Liu$^{2}$, Hui Hu$^{2}$, and Han Pu$^{1}$}

\affiliation{$^{1}$Department of Physics and Astronomy, and Rice Quantum Institute,
Rice University, Houston, TX 77251, USA \\
$^{2}$ARC Centre of Excellence for Quantum-Atom Optics, Centre for
Atom Optics and Ultrafast Spectroscopy, Swinburne University of Technology,
Melbourne 3122, Australia }

\date{\today}
\begin{abstract}
We investigate theoretically BEC-BCS crossover physics in the presence of a Rashba spin-orbit
coupling in a system of two-component Fermi gas with and without a Zeeman field that breaks the population balance between the two components. A new bound state (Rashba pair) emerges because of the spin-orbit interaction. We study
the properties of Rashba pairs using a standard pair fluctuation theory. At zero temperature, the
Rashba pairs condense into a macroscopic mixed spin state. We discuss in detail the experimental
signatures for observing the condensation of Rashba pairs by calculating various physical observables which characterize the properties of the system and can be measured in experiment.
\end{abstract}


\maketitle

\section{Introduction}
Since its recent realization in cold atomic systems \cite{lin1,lin2,lin3,lin4,zhang}, artificial gauge field has received tremendous attention. The concept of gauge field is ubiquitous, a classical example of which is electromagnetism. The achievement of the above mentioned experiments allows us to simulate charged particles moving in electromagnetic fields using neutral atoms. The more recent realization of a non-Abelian gauge field in a system of $^{87}$Rb condensate \cite{lin4} provides us a system of spinor quantum gas whose internal (pseudo-)spin degrees of freedom and external spatial degrees of freedom are intimately coupled. Novel quantum states will emerge in such spin-orbit coupled systems \cite{dalibard}. Although experiments on artificial gauge field have so far only been carried out in bosonic systems, we have no reason to doubt that they will soon be extended to fermionic systems. Theoretically, there have been a number of papers focusing on the interesting properties of spin-orbit coupled Fermi gases in both three dimension (3D) \cite{shenoy,shenoy1,hu,zhai,chuanwei,chuanwei1,yi,sau,iskin,iskin1,baym,melo,melo1} and 2D \cite{two1,two2,two3,two4}, most of the them appeared just over the past few months.

The salient features of spin-orbit coupled fermions include: enhanced pairing field \cite{shenoy1,hu,chuanwei1}, mixed spin pairing \cite{gorkov}, non-trivial topological order \cite{chuanwei1,tewari}, and possible existence of Majorana fermion \cite{sato}, etc. The purpose of the present paper is to extend our earlier work \cite{hu} by providing a detailed description of the theoretical techniques and by including the effect of a Zeeman field which not only breaks the population balance, but also may induce topological phase  transitions in the system. We start from a discussion of the two-body problem, followed by a detailed study of the many-body system. We present our calculations of various important physical observables such as the single-particle spectrum, density of states, spin structure factors, etc., which may be used to characterize the system experimentally.

The paper is organized as follows. The model Hamiltonian and the general technique are presented in Sec.~\ref{formalism}. The results for the two-body and the many-body problem are presented in Sec.~\ref{2body} and \ref{mbody}, respectively. Finally we provide an outlook and conclusion in Sec.~\ref{oc}.

\section{Physical model and general technique}
\label{formalism}
In this section, we first present the model Hamiltonian of interest and then give a detailed description of the functional path integral formalism employed in deriving the relevant equations. We choose this formalism as it allows us to present a unified treatment for both the two-body and the many-body physics.

\subsection{Model Hamiltonian}

Here we consider the BEC-BCS crossover theory in the presence of a Rashba
spin-orbit (SO) coupling $\lambda (\hat{k}_{y}\hat{\sigma}_{x}-\hat{k}_{x}\hat{\sigma}%
_{y})$ in the $x$-$y$ plane, together with an external Zeeman field
$h\hat{\sigma}_{z}$. The Zeeman field acts to break the population
balance between the two spin components of the fermions. The
second-quantized Hamiltonian for a uniform system reads,
\begin{eqnarray}
\mathcal{H} &=& \int d\mathbf{r}\left\{ \psi ^{+}\left[ \xi _{\mathbf{k}}+h\hat{%
\sigma}_{z}+\lambda
(\hat{k}_{y}\hat{\sigma}_{x}-\hat{k}_{x}\hat{\sigma}_{y})\right]
\psi \right. \nonumber \\
&& \left. +U_{0}\psi _{\uparrow }^{+}\left( \mathbf{r}\right) \psi _{\downarrow
}^{+}\left( \mathbf{r}\right) \psi _{\downarrow }\left( \mathbf{r}\right)
\psi _{\uparrow }\left( \mathbf{r}\right) \right\} , \label{H}
\end{eqnarray}%
where $\xi _{\mathbf{k}}=\hbar ^{2}\hat{k}^{2}/(2m)-\mu $ with $\mu$ being the chemical potential, and $\psi \left( \mathbf{%
r}\right) =[\psi _{\uparrow }\left( \mathbf{r}\right) ,\psi
_{\downarrow }\left( \mathbf{r}\right) ]^T$, $\psi _{\sigma }\left(
\mathbf{r}\right) $ is the fermionic annihilation operator for
spin-$\sigma$ atom. Here $h$ is the strength of the Zeeman field and
$\lambda $ is the Rashba SO coupling constant. Without loss of
generality, we take both $h$ and $\lambda$ to be non-negative. The
last term in Eq.~(\ref{H}) represents the two-body contact $s$-wave
interaction between un-like spins.

The SO coupling term has already some interesting effect on the single-particle physics. The single-particle spectrum (i.e., the eigenenergy of the dressed states) can be straightforwardly obtained as
\begin{equation}
\epsilon_{\bf k, \pm} = \xi_{\bf k} \pm \sqrt{h^2 + \lambda^2 k_{\perp}^2} \,,\label{spe} \end{equation}
where $k_\perp = \sqrt{k_x^2+k_y^2}$ is the magnitude of the transverse momentum. The lowest single-particle state occurs at $k_z=0$ and
\begin{equation}
k_\perp = \left\{ \begin{array}{cc} \sqrt{m^2 \lambda^2/\hbar^4 -h^2/\lambda^2} \,, & h < m\lambda^2/\hbar^2 \\ 0 \,, & {\rm otherwise} \end{array} \right. \,, \label{h}
\end{equation}
with the corresponding lowest single-particle energy as (taking $\mu=0$):
\begin{equation}
\epsilon_{\rm min} = \left\{ \begin{array}{cc} {-m \lambda^2/(2\hbar^2) -\hbar^2 h^2/(2m\lambda^2)} \,, & h < m\lambda^2/\hbar^2 \\ -h \,, & {\rm otherwise} \end{array} \right. \,. \label{min}
\end{equation}
Hence for $h$ smaller than a threshold value $m\lambda^2/\hbar^2$,
the single-particle ground state is infinitely degenerate and occurs
along a ring in momentum space centered at ${\bf k}=0$ and lies in
the transverse plane. The radius of the `Rashba ring' decreases as
$h$ is increased and vanishes when $h$ exceeds the threshold value,
in which case the ground state becomes non-degenerate and occurs at
momentum ${\bf k}=0$.

\subsection{Functional Path Integral Formalism}

We employ the functional path integral method \cite{sademelo,hu2} to
study the problem and consider the partition function,
\begin{equation}
\mathcal{Z}=\int \mathcal{D}[\psi \left( \mathbf{r},\tau \right) ,\bar{\psi}%
\left( \mathbf{r},\tau \right) ]\exp \left\{ -S\left[ \psi \left( \mathbf{r}%
,\tau \right) ,\bar{\psi}\left( \mathbf{r},\tau \right) \right] \right\} ,
\end{equation}%
where the action
\[ S\left[ \psi ,\bar{\psi}\right] =\int_{0}^{\beta }d\tau \left[ \int d%
\mathbf{r}\sum_{\sigma }\bar{\psi}_{\sigma }\left( \mathbf{r},\tau
\right)
\partial _{\tau }\psi _{\sigma }\left( \mathbf{r},\tau \right) +\mathcal{H}\left(
\psi ,\bar{\psi}\right) \right] \,.\] is written as an integral over
imaginary time $\tau$.
Here $\beta =1/(k_{B}T)$ is the inverse temperature and $\mathcal{H}%
\left( \psi ,\bar{\psi}\right) $ is obtained by replacing the field
operators $\psi ^{+}$ and $\psi $ with the grassmann variables $\bar{\psi}$
and $\psi $, respectively. We can use the Hubbard-Stratonovich
transformation to transform the quartic interaction term into the quadratic form as:
\begin{widetext}
\begin{equation}
e^{-U_{0}\int dxd\tau \bar{\psi}_{\uparrow }\bar{\psi}_{\downarrow }\psi
_{\downarrow }\psi _{\uparrow }}=\int \mathcal{D}\left[ \Delta ,\bar{\Delta}%
\right]\, \exp \left\{ \int_{0}^{\beta }d\tau \int d\mathbf{r}\left[ \frac{%
\left\vert \Delta \left( \mathbf{r},\tau \right) \right\vert ^{2}}{U_{0}}%
+\left( \bar{\Delta}\psi _{\downarrow }\psi _{\uparrow }\mathbf{+}\Delta
\bar{\psi}_{\uparrow }\bar{\psi}_{\downarrow }\right) \right] \right\} \,,
\end{equation}%
from which the pairing field $\Delta \left( \mathbf{r},\tau \right) $ is defined.

Let us now formally introduce the 4-dimensional Nambu spinor $\Phi \left(
\mathbf{r,}\tau \right) \equiv \lbrack \psi _{\uparrow },\psi _{\downarrow }%
\mathbf{,}\bar{\psi}_{\uparrow },\bar{\psi}_{\downarrow }]^T$ and rewrite the
action as,
\begin{equation}
\mathcal{Z}=\int \mathcal{D}[\Phi ,\bar{\Phi}\mathbf{;}\,\Delta ,\bar{\Delta}%
]\, \exp \left\{ -\int \! d\tau \! \! \int \! d\mathbf{r} \! \! \int \! d\tau' \! \! \int \! d\mathbf{r'}\left[ -\frac{1}{2}\bar{%
\Phi}(\mathbf{r},\tau) \mathcal{G}^{-1}
\Phi(\mathbf{r'},\tau') -\frac{\left\vert \Delta \left(
\mathbf{r},\tau \right)  \right\vert ^{2}}{U_{0}} \delta(\mathbf{r}-\mathbf{r'})\delta(\tau-\tau') \right] - \frac{\beta}{V} \sum_{\mathbf{%
k}}\xi _{\mathbf{k}}\right\} ,
\end{equation}%
where $V$ is the quantization volume and the single-particle Green function is given by,
\begin{equation}
\mathcal{G}^{-1}=\left[
\begin{array}{cc}
-\partial _{\tau }-\xi _{\mathbf{k}}-h\hat{\sigma}_{z}-\lambda (k_{y}\hat{%
\sigma}_{x}-k_{x}\hat{\sigma}_{y}) & i\Delta \hat{\sigma}_{y} \\
-i\bar{\Delta}\hat{\sigma}_{y} & -\partial _{\tau }+\xi _{\mathbf{k}}+h\hat{%
\sigma}_{z}-\lambda (k_{y}\hat{\sigma}_{x}+k_{x}\hat{\sigma}_{y})%
\end{array}%
\right]\delta(\mathbf{r}-\mathbf{r'})\delta(\tau-\tau') \,,
\label{green}
\end{equation}%
\end{widetext}
where the Pauli matrix $\hat{\sigma}_{i}$ ($i=0,x,y,z$) describes the spin
degrees of freedom. Here the momentum $k_{\alpha }$ ($\alpha =x,y,z$) should
be regarded as the operators in real space.

Integrating out the original fermionic fields, we may rewrite the partition function as
\[ \mathcal{Z}=\int \mathcal{D}[\Delta ,\bar{\Delta}] \, \exp \left\{
-S_{\rm eff}\left[ \Delta ,\bar{\Delta}\right] \right\} \,,\] where the effective action is given by
\begin{eqnarray*}
S_{\rm eff}\left[ \Delta ,\bar{\Delta}\right] &=& \int_{0}^{\beta }d\tau \int d%
\mathbf{r}\left\{ -\frac{\left\vert \Delta \left( \mathbf{r},\tau \right)
\right\vert ^{2}}{U_{0}}\right\}  \\ && \; -\frac{1}{2}\text{Tr}\ln \left[ -\mathcal{G}%
^{-1}\right] + \frac{\beta}{V} \sum_{\mathbf{k}}\xi _{\mathbf{k}}.
\end{eqnarray*}%
where
the trace is over all the spin, spatial, and temporal degrees of freedom. To
proceed, we restrict to the gaussian fluctuation and expand $\Delta \left(
\mathbf{r},\tau \right) =\Delta _{0}+\delta \Delta \left( \mathbf{r},\tau
\right) $. The effective action is then decomposed accordingly as $S_{\rm eff}=S_{0}+%
\Delta S$, where the saddle-point action is
\begin{equation}
\label{S0}
S_{0} =\int_{0}^{\beta }d\tau \int d\mathbf{r}\left( -\frac{\Delta _{0}^{2}%
}{U_{0}}\right) -\frac{1}{2}\text{Tr}\ln \left[ -\mathcal{G}_{0}^{-1}\right]
+\frac{\beta}{V} \sum_{\mathbf{k}}\xi _{\mathbf{k}} \,,
\end{equation}
where $\mathcal{G}_{0}^{-1}$ has the same form as $\mathcal{G}^{-1}$ in Eq.~(\ref{green}) with $\Delta$ replaced by $\Delta_0$, and the fluctuating action takes the form
\[ \Delta S =\int_{0}^{\beta }d\tau \int d\mathbf{r}\left\{ -\frac{\left\vert
\delta \Delta \left( \mathbf{r},\tau \right) \right\vert ^{2}}{U_{0}}+\frac{1%
}{2}\left( \frac{1}{2}\right) \text{Tr}\left( \mathcal{G}_{0}\Sigma \right)
^{2}\right\} \,, \]
with
\begin{equation}
\Sigma =\left(
\begin{array}{cc}
0 & i\delta \Delta \hat{\sigma}_{y} \\
-i\delta \bar{\Delta}\hat{\sigma}_{y} & 0%
\end{array}%
\right) .
\end{equation}
being the self energy.

\subsection{Vertex Function}
The low-energy effective two-body interaction is characterized by the vertex function, which we derive in this section.
We shall consider the normal state where the pairing field vanishes, i.e., $\Delta _{0}=0$. In this case, the inverse Green function $\mathcal{G}_0^{-1}$ has a diagonal form and can be easily inverted to give :
\begin{equation}
\mathcal{G}_{0}\left( { k}\right) =\left(
\begin{array}{cc}
\hat{g}_{+}({ k}) & 0 \\
0 & \hat{g}_{-}({ k})%
\end{array}%
\right) ,
\end{equation}%
where $k \equiv ({\bf k}, i\omega_m)$ and
\begin{widetext}
\begin{eqnarray}
\hat{g}_{+}(k) &=&\frac{1}{i\omega _{m}-\xi _{\mathbf{k}}-h\hat{\sigma}%
_{z}-\lambda (k_{y}\hat{\sigma}_{x}-k_{x}\hat{\sigma}_{y})}=\frac{i\omega
_{m}-\xi _{\mathbf{k}}+h\hat{\sigma}_{z}+\lambda (k_{y}\hat{\sigma}_{x}-k_{x}%
\hat{\sigma}_{y})}{\left( i\omega _{m}-\xi _{\mathbf{k}}\right) ^{2}-\left[
h^{2}+\lambda ^{2}\left( k_{x}^{2}+k_{y}^{2}\right) \right] }, \\
\hat{g}_{-}({ k}) &=&\frac{1}{i\omega _{m}+\xi _{\mathbf{k}}+h\hat{\sigma}%
_{z}-\lambda (k_{y}\hat{\sigma}_{x}+k_{x}\hat{\sigma}_{y})}=\frac{i\omega
_{m}+\xi _{\mathbf{k}}-h\hat{\sigma}_{z}+\lambda (k_{y}\hat{\sigma}_{x}+k_{x}%
\hat{\sigma}_{y})}{\left( i\omega _{m}+\xi _{\mathbf{k}}\right) ^{2}-\left[
h^{2}+\lambda ^{2}\left( k_{x}^{2}+k_{y}^{2}\right) \right] }.
\end{eqnarray}%
\end{widetext}
After some algebra, we may obtain the fluctuating part of the action as
\begin{equation}
\Delta S=k_{B}T\frac{1}{V}\sum_{q=\mathbf{q},i\nu _{n}}\left[ -\Gamma ^{-1}\left(
q\right) \right] \delta \Delta (q)\delta \bar{\Delta}(q) \,,
\end{equation}%
where the inverse vertex function is given by
\begin{widetext}
\begin{equation}
\Gamma ^{-1}\left( q\right) =\frac{1}{U_{0}}+k_{B}T\frac{1}{V}\sum_{\mathbf{k},i\omega
_{m}}\left[ \frac{1/2}{\left( i\omega _{m}-\epsilon_{\mathbf{k},+}\right) \left(
i\nu _{n}-i\omega _{m}-\epsilon_{\mathbf{q}-\mathbf{k},+}\right) }+\frac{1/2}{%
\left( i\omega _{m}-\epsilon_{\mathbf{k},-}\right) \left( i\nu _{n}-i\omega _{m}-\epsilon_{%
\mathbf{q}-\mathbf{k},-}\right) }-A_{res}\right] ,\label{gamma}
\end{equation}%
where $\epsilon_{\mathbf{k},\pm}$ are the single-particle spectrum in Eq.~(\ref{spe}) and
\begin{equation}
A_{res}=\frac{\sqrt{h^{2}+\lambda ^{2}k_{\perp }^{2}}\sqrt{h^{2}+\lambda
^{2}\left( \mathbf{q}-\mathbf{k}\right) _{\perp }^{2}}+h^{2}+\lambda
^{2}k_{x}\left( q_{x}-k_{x}\right) +\lambda ^{2}k_{y}\left(
q_{y}-k_{y}\right) }{\left( i\omega _{m}-\epsilon_{\mathbf{k},+}\right) \left(
i\omega _{m}-\epsilon_{\mathbf{k},-}\right) \left( i\nu _{n}-i\omega _{m}-\epsilon_{%
\mathbf{q}-\mathbf{k},+}\right) \left( i\nu _{n}-i\omega _{m}-\epsilon_{\mathbf{q}-%
\mathbf{k},-}\right) }.
\end{equation}%
The summation over $i\omega _{m}$ in Eq.~(\ref{gamma}) can be done explicitly, after which we find that,
\begin{eqnarray}
\Gamma ^{-1}\left( q\right) &=&\frac{m}{4\pi \hbar ^{2}a_{s}}+\frac{1}{2V}%
\sum_{\mathbf{k}}\left[ \frac{f\left( \epsilon_{\mathbf{q}/2+\mathbf{k},+}\right)
+f\left( \epsilon_{\mathbf{q}/2-\mathbf{k},+}\right) -1}{i\nu _{n}-\epsilon_{\mathbf{q}/2+%
\mathbf{k},+}-\epsilon_{\mathbf{q}/2-\mathbf{k},+}}+\frac{f\left( \epsilon_{\mathbf{q}/2+%
\mathbf{k},-}\right) +f\left( \epsilon_{\mathbf{q}/2-\mathbf{k},-}\right) -1}{i\nu
_{n}-\epsilon_{\mathbf{q}/2+\mathbf{k},-}-\epsilon_{\mathbf{q}/2-\mathbf{k},-}}-\frac{1}{%
\epsilon _{\mathbf{k}}}\right]  \notag \\
&&-\frac{1}{4V}\sum_{\mathbf{k}}\left[ 1+\frac{h^{2}+\lambda ^{2}\left(
q_{\perp }^{2}/4-k_{\perp }^{2}\right) }{\sqrt{h^{2}+\lambda ^{2}\left(
\mathbf{q}/2+\mathbf{k}\right) _{\perp }^{2}}\sqrt{h^{2}+\lambda ^{2}\left(
\mathbf{q}/2-\mathbf{k}\right) _{\perp }^{2}}}\right] C_{res}(\mathbf{q}%
,i\nu _{n};\mathbf{k}), \label{gamma1}
\end{eqnarray}%
where $f(x) =1/(e^{\beta x}+1)$ is the Fermi distribution function and
\begin{eqnarray}
C_{res} &=&+\frac{\left[ f\left( \epsilon_{\mathbf{q}/2+\mathbf{k},+}\right)
+f\left( \epsilon_{\mathbf{q}/2-\mathbf{k},+}\right) -1\right] }{i\nu _{n}-\epsilon_{%
\mathbf{q}/2+\mathbf{k},+}-\epsilon_{\mathbf{q}/2-\mathbf{k},+}}+\frac{\left[
f\left( \epsilon_{\mathbf{q}/2+\mathbf{k},-}\right) +f\left( \epsilon_{\mathbf{q}/2-%
\mathbf{k},-}\right) -1\right] }{i\nu _{n}-\epsilon_{\mathbf{q}/2+\mathbf{k},-}-\epsilon_{%
\mathbf{q}/2-\mathbf{k},-}}  \notag \\
&&-\frac{\left[ f\left( \epsilon_{\mathbf{q}/2+\mathbf{k},+}\right) +f\left( \epsilon_{%
\mathbf{q}/2-\mathbf{k},-}\right) -1\right] }{i\nu _{n}-\epsilon_{\mathbf{q}/2+%
\mathbf{k},+}-\epsilon_{\mathbf{q}/2-\mathbf{k},-}}-\frac{\left[ f\left( \epsilon_{\mathbf{%
q}/2+\mathbf{k},-}\right) +f\left( \epsilon_{\mathbf{q}/2-\mathbf{k},+}\right) -1%
\right] }{i\nu _{n}-\epsilon_{\mathbf{q}/2+\mathbf{k},-}-\epsilon_{\mathbf{q}/2-\mathbf{k}%
,+}}.
\end{eqnarray}%
\end{widetext}
In writing the above equations, we have replaced the bare interaction strength $U_0$ in favor of the $s$-wave scattering length $a_s$ using \[ \frac{1}{U_0} = \frac{m}{4\pi \hbar^2 a_s} -\frac{1}{V} \sum_{\bf k} \frac{1}{2\epsilon_{\bf k}} \]with $\epsilon_{\bf k} = \hbar^2 k^2/(2m)$.

\section{Results on Two-body problem}
\label{2body}

Let us first consider the two-body problem. The corresponding two-body inverse vertex function can be obtained from Eq.~(\ref{gamma1}) by discarding the Fermi
distribution function and by setting chemical potential $\mu =0$. This leads to
\begin{widetext}
\begin{eqnarray}
\Gamma _{\rm 2b}^{-1}\left( q\right) &=&\frac{m}{4\pi \hbar ^{2}a_{s}}-\frac{1}{2V%
}\sum_{\mathbf{k}}\left[ \frac{1}{i\nu _{n}-\epsilon_{\mathbf{q}/2+\mathbf{k},+}-\epsilon_{%
\mathbf{q}/2-\mathbf{k},+}}+\frac{1}{i\nu _{n}-\epsilon_{\mathbf{q}/2+\mathbf{k}%
,-}-\epsilon_{\mathbf{q}/2-\mathbf{k},-}}+\frac{1}{\epsilon _{\mathbf{k}}}\right]
\notag \\
&&+\frac{1}{4V}\sum_{\mathbf{k}}\left[ 1+\frac{h^{2}+\lambda ^{2}\left(
q_{\perp }^{2}/4-k_{\perp }^{2}\right) }{\sqrt{h^{2}+\lambda ^{2}\left(
\mathbf{q}/2+\mathbf{k}\right) _{\perp }^{2}}\sqrt{h^{2}+\lambda ^{2}\left(
\mathbf{q}/2-\mathbf{k}\right) _{\perp }^{2}}}\right] \bar{C}_{res}^{2b}(%
\mathbf{q},i\nu _{n};\mathbf{k}),
\end{eqnarray}%
where
\begin{eqnarray}
\bar{C}_{res}^{2b} &=&+\frac{1}{i\nu _{n}-\epsilon_{\mathbf{q}/2+\mathbf{k},+}-\epsilon_{%
\mathbf{q}/2-\mathbf{k},+}}+\frac{1}{i\nu _{n}-\epsilon_{\mathbf{q}/2+\mathbf{k}%
,-}-\epsilon_{\mathbf{q}/2-\mathbf{k},-}}  \notag \\
&&-\frac{1}{i\nu _{n}-\epsilon_{\mathbf{q}/2+\mathbf{k},+}-\epsilon_{\mathbf{q}/2-\mathbf{k%
},-}}-\frac{1}{i\nu _{n}-\epsilon_{\mathbf{q}/2+\mathbf{k},-}-\epsilon_{\mathbf{q}/2-%
\mathbf{k},+}}.
\end{eqnarray}%

One important question concerning the two-body system is whether there exist bound states. The zero-momentum bound state energy $E_B$ can be determined from the vertex function using the following relation ($i\nu
_{n}\rightarrow \omega +i0^{+}$):
\begin{equation}
{\rm Re} \left[ \Gamma _{\rm 2b}^{-1}\left( \mathbf{q}=0;\omega =E_{B}\right) =0 \right] \,,
\end{equation}%
from which we may derive the equation for the bound state energy as
\begin{equation}
\frac{m}{4\pi \hbar ^{2}a_{s}}-\frac{1}{2V}\sum_{\mathbf{k}}\left[ \frac{1}{%
E_{B}-2\epsilon_{\mathbf{k},+}}+\frac{1}{E_{B}-2\epsilon_{\mathbf{k},-}}+\frac{1}{\epsilon
_{\mathbf{k}}}\right] +\frac{1}{V}\sum_{\mathbf{k}}\frac{4h^{2}}{\left( E_{B}-2\epsilon
_{\mathbf{k}}\right) \left( E_{B}-2\epsilon_{\mathbf{k},+}\right) \left( E_{B}-2\epsilon_{%
\mathbf{k},-}\right) }=0\,.  \label{Eb}
\end{equation}%
\end{widetext}
A bound state exists if its energy satisfies \[E_B < 2 \epsilon_{\rm
min} \,,\] where $\epsilon_{\rm min}$ is the lowest single-particle
energy defined in Eq.~(\ref{min}).

\begin{figure}[tbh]
\includegraphics[width=8cm]{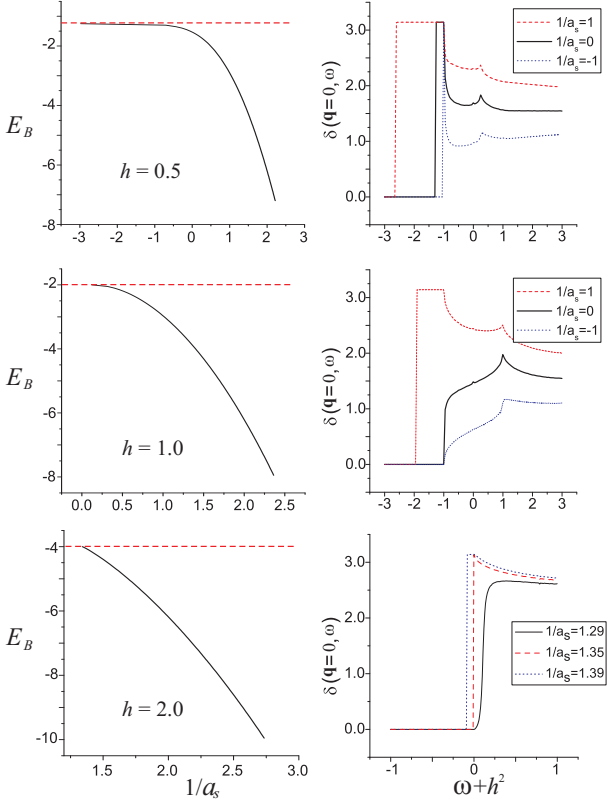}\caption{(color online). Left panel: Bound state energies $E_B$ as functions of scattering length for
different Zeeman field $h$. The horizontal dashed lines represent the threshold energy $2\epsilon_{\rm min}$. Right panel: Corresponding two-body phase shifts
$\delta ({\bf q}=0, \omega)$ for different Zeeman field $h$. $E_B$
and $h$ are in units of $m \lambda^2/\hbar^2$, $a_s$ is in units of
$\hbar^2 /(m\lambda)$. }
\label{bound state}%
\end{figure}

We solve Eq.~(\ref{Eb}) numerically to find $E_B$ and the results
are shown in the left panel of Fig.~\ref{bound state}, where we
plot $E_B$ as a function of the contact interaction strength for three
different values of the Zeeman field ($h=0.5$, 1.0 and $2.0 \, m\lambda ^{2}/\hbar
^{2}$). For $h<m\lambda^2/\hbar^2$, i.e., when the Rashba ring
exists as the lowest single-particle state [see Eq.~(\ref{h})], we
always find one bound state solution regardless of the sign of
$a_s$. It is well known that in the absence of the SO coupling,
two-body bound state does not exist on the BCS side (i.e., $a_s<0$).
The existence of the Rashba ring induced by the SO coupling enhances
the density of states near the single-particle ground state and
favors the formation of bound state \cite{shenoy}. By contrast, for
$h \ge m\lambda^2/\hbar^2$, the Rashba ring collapses to a point and
two-body bound state only occurs on the BEC side. Furthermore, the larger the $h$ is, the stronger attractive interaction (i.e., larger $a_s^{-1}$) is required to have a bound state. For example, at $h=1.0 \, m\lambda ^{2}/\hbar
^{2}$, the bound state exists for $a_s^{-1}>0$; at $h=2.0 \, m\lambda ^{2}/\hbar
^{2}$, the bound state exist only for $a_s^{-1} > 1.35 \, m\lambda/\hbar^2$ (see Fig.~\ref{bound state}).


If there is a tightly bound state, the inverse vertex function is
simply the Green function of the bound pair. A bound state can
therefore be examined by calculating the phase shift \cite{nsr,Liu}
\begin{equation}
\label{ps}
\delta({\bf q},\omega)=-\mathop{\rm Im} \left\{ \ln[-\Gamma_{\rm
2b}^{-1}({\bf q},i\nu_{n}\rightarrow\omega+i0^{+})] \right\}\,.
\end{equation} 
In
the right panel of Fig.~\ref{bound state}, we display the phase
shift at ${\bf q}=0$. When a bound state occurs, the phase shift
will have a discontinuous jump of $\pi$ at the corresponding energy
as can be seen in the figure. 
The phase shift calculation via Eq.~(\ref{ps}) and the bound state energy calculation via Eq.~(\ref{Eb}) thus corroborate with each other.

\begin{figure}[tbh]
\includegraphics[width=6.5cm]{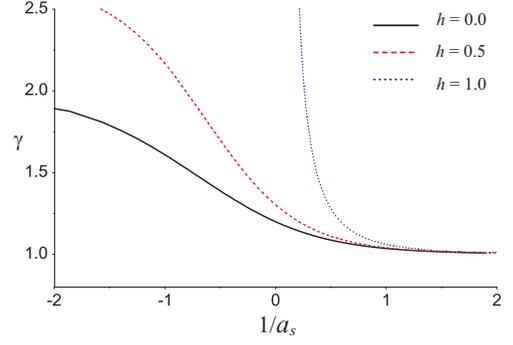}\caption{(color online). Effective mass $\gamma \equiv M_\perp/(2m)$ as functions of scattering length for
different Zeeman field $h$. $h$ is in units of $m \lambda^2/\hbar^2$, and $a_s$ is in units of $\hbar^2 /(m\lambda)$. For $h \ge 1$, the two-body bound state only exists for $a_s>0$.}
\label{mass}%
\end{figure}

An important quantity that characterizes the properties of the bound state is its effective mass. At small momentum $|{\bf q}|$, we
may assume that the bound state has a well-defined dispersion, $\epsilon
_{\bf q}^B=\hbar ^2q_{\perp }^2/(2M_\perp)+\hbar ^2q_z^2/(4m)$, where $M_\perp$ is the effective mass in the transverse plane. Due to the nature of the Rashba SO coupling, the effective mass of the bound state along the $z$-axis is simple twice of the atomic mass and is not affected by the spin-orbit term. For a given ${\bf q}$, we determine $\epsilon_{\bf q}^B$ from the equation:
$
\mathop{\rm Re}
\left[ \Gamma _{\rm 2b}^{-1}\left( {\bf q};\omega =E_B+\epsilon _{\bf q}^B\right) =0 \right] \,.
$
By Taylor expanding the two-body inverse vertex function around $|{\bf q}|=0$, and after some tedious but straightforward calculation, we obtain:
\begin{equation}
\frac{1}{\gamma} \equiv \frac{2m}{M_\perp} = 1- \frac{4m\lambda^2}{\hbar^2} \frac{Y}{X}\,,
\end{equation}
where

\begin{widetext}
\begin{eqnarray*}
X &=& \frac{1}{V}\sum_{{\bf k}}\left\{ \left[ \frac{2h^{2}}{h^{2}+\lambda ^{2}k_{\perp }^{2}%
}\right] \frac{1}{\left( E_{B}-2\epsilon _{{\bf k}}\right) ^{2}}+\frac{%
\lambda ^{2}k_{\perp }^{2}}{h^{2}+\lambda ^{2}k_{\perp }^{2}}\left[ \frac{1}{%
\left( E_{B}-2\epsilon_{{\bf k},+}\right) ^{2}}+\frac{1}{\left( E_{B}-2\epsilon_{{\bf k}%
,-}\right) ^{2}}\right] \right\} ,\\
Y &=&\frac{1}{V}\sum_{{\bf k}}\left\{ \frac{\lambda ^{2}k_{\perp
}^{2}\left( 3h^{2}+\lambda ^{2}k_{\perp }^{2}\right) }{\left(
h^{2}+\lambda ^{2}k_{\perp }^{2}\right) ^{2}}\frac{1}{ \left(
E_{B}-2\epsilon_{{\bf k},+}\right) \left( E_{B}-2\epsilon_{{\bf
k},-}\right) } \frac{1}{\left( E_{B}-2\epsilon _{{\bf k}}\right)}
-\frac{h^{2}\lambda
^{2}k_{\perp }^{2}}{\left( h^{2}+\lambda ^{2}k_{\perp }^{2}\right) ^{2}}%
\frac{1}{\left( E_{B}-2\epsilon _{{\bf k}}\right) ^{3}}  \right. \\
&& \;\; \left.  -\frac{\lambda ^{2}k_{\perp }^{2}(2h^{2}+\lambda ^{2}k_{\perp }^{2})%
}{\left( h^{2}+\lambda ^{2}k_{\perp }^{2}\right) ^{2}}\left[
\frac{\left( E_{B}-2\epsilon _{{\bf k}}\right) }{\left(
E_{B}-2\epsilon_{{\bf k},+}\right) ^{2}\left( E_{B}-2\epsilon_{{\bf
k},-}\right) ^{2}}\right] \right\} .
\end{eqnarray*}%

Figure \ref{mass} displays the effective mass $M_\perp$ as functions of the scattering length $a_s$ for several values of the Zeeman field strength ($h$). $M_\perp$ monotonically decreases as $1/a_s$ increases. In the BEC limit where $a_s \rightarrow 0^+$, $M_\perp \rightarrow 2m$ independent of the value of $h$.


\section{Results on Many-body problem}
\label{mbody} We now turn to the discussion of the many-body
properties. In this work, we only consider the mean-field properties
of the system, while the effect of fluctuations will be studied in
the future. In the mean-field level, the order parameter $\Delta_0$
is a constant and the corresponding momentum space
single-particle Green function takes the form ($k=\mathbf{k}%
,i\omega _{m}$),
\begin{equation}
\mathcal{G}_{0}^{-1}\left( k\right) =\left[
\begin{array}{cc}
i\omega _{m}-\xi _{\mathbf{k}}-h\hat{\sigma}_{z}-\lambda (k_{y}\hat{\sigma}%
_{x}-k_{x}\hat{\sigma}_{y}) & i\Delta _{0}\hat{\sigma}_{y} \\
-i\Delta _{0}\hat{\sigma}_{y} & i\omega _{m}+\xi _{\mathbf{k}}+h\hat{\sigma}%
_{z}-\lambda (k_{y}\hat{\sigma}_{x}+k_{x}\hat{\sigma}_{y})%
\end{array}%
\right]\,.  \label{gf0}
\end{equation}
Plugging this into Eq.~(\ref{S0}), we may obtain the
thermodynamic potential as
\begin{equation}
\Omega _{0}=\frac{1}{V}\sum_{\mathbf{k}}\left( \xi _{\mathbf{k}}-\frac{E_{\mathbf{k+}%
}+E_{\mathbf{k-}}}{2}\right) -\frac{\Delta _{0}^{2}}{U_{0}}-k_{B}T\sum_{%
\mathbf{k},\alpha =\pm }\ln \left[ 1+e^{-E_{\mathbf{k}\alpha }/k_{B}T}\right]
\,.
\end{equation}%
where $E_{{\bf k}\pm }=\sqrt{\xi _{\bf k}^{2}+\Delta _{0}^{2}+h^{2}+\lambda
^{2}k_{\perp }^{2}\pm 2\sqrt{(h^{2}+\lambda ^{2}k_{\perp }^{2})\xi
_{\bf k}^{2}+h^{2}\Delta _{0}^{2}}}$ is the quasi-particle dispersion. The chemical potential and order parameter
should be determined by,
\begin{eqnarray*}
0 =\frac{\partial \Omega _{0}}{\partial \Delta _{0}}\,,\quad
N &=&-\frac{\partial \Omega _{0}}{\partial \mu }\,,
\end{eqnarray*}%
from which we derive the gap and the number equations as follows:
\begin{eqnarray}
\frac{1}{U_{0}} &=&\frac{1}{V}\underset{{\bf k},\alpha=\pm }{\sum } \frac{2f(E_{{\bf k}\alpha})-1}{%
4E_{{\bf k}\alpha}}\left[1 + \alpha \frac{h^{2}}{\sqrt{(h^{2}+\lambda ^{2}k_{\perp }^{2})\xi
_{\bf k}^{2}+h^{2}\Delta _{0}^{2}}} \right] \,, \label{gap}\\
n &=& \frac{N}{V}=\frac{1}{V}\underset{\bf k}{\sum } \left\{ 1+\sum_{\alpha = \pm} \frac{\xi _{\bf k}[2f(E_{{\bf k}\alpha})-1]}{2E_{{\bf k}\alpha}}%
\left[1+ \alpha \frac{h^{2}+\lambda ^{2}k_{\perp }^{2}}{\sqrt{(h^{2}+\lambda ^{2}k_{\perp
}^{2})\xi _{\bf k}^{2}+h^{2}\Delta _{0}^{2}}} \right] \right\}\,. \label{number}
\end{eqnarray}
\end{widetext}

In the following, we will show various quantities of physical
interest, obtained from solving Eqs.~(\ref{gap}) and (\ref{number})
self-consistently. We focus on the zero-temperature case, although
Eqs.~(\ref{gap}) and (\ref{number}) are valid for finite
temperatures.

\begin{figure}[tbh]
\includegraphics[width=8.5cm]{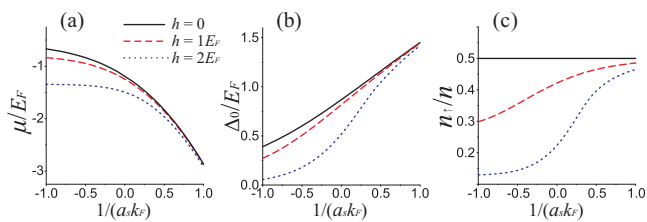}
\caption{(color online). Chemical potential $\mu $ (a), pairing gap $\Delta _{0}$ (b), and population of spin-up component $n_\uparrow$ (c) as functions of scattering length $a_s$ for different values of the Zeeman field $h$ at $\lambda k_F/E_F=2$. Here $k_F=(3\pi^2n)^{1/3}$ and $E_F=\hbar^2 k_F^2/(2m)$ are the Fermi momentum and Fermi energy, respectively.}%
\label{chemical potential and gap}%
\end{figure}

We show in Fig.~\ref{chemical potential and gap}(a) and (b) the chemical potential $\mu $ and the
pairing gap $\Delta _{0}$, respectively, as functions of the scattering length for different values of the Zeeman field. The Zeeman field tends to suppress the pairing gap. In Fig.~\ref{chemical potential and gap}(c), we plot the population of the spin-up component. For zero Zeeman field, we always have equal population in both spin components. For $h>0$, spin-up component has less population. In all cases, the effect of the Zeeman field reduces as the BEC limit (i.e., $1/a_s \rightarrow + \infty$) is approached.

\begin{figure}[tbh]
\includegraphics[width=8cm]{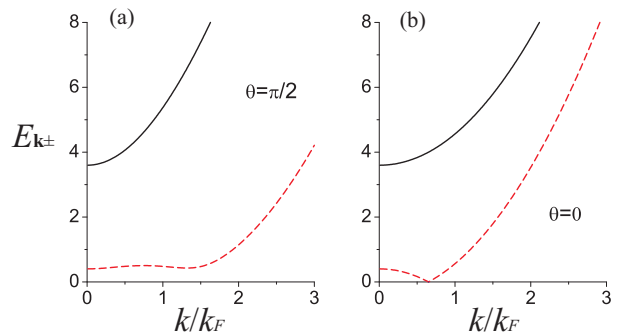}
\caption{(color online). Quasi-particle dispersion spectrum $E_{{\bf k}+}$ (solid lines) and $E_{{\bf k}-}$ (dashed lines) shown in in units of $E_F$ for ${\bf k}$ in the transverse plane (a, $\theta=\pi/2$) and along the $z$-axis (b, $\theta=0$) for $%
\protect\lambda k_{F}/E_{F}=2$, $h/E_{F}=2$. }%
\label{spectrum}%
\end{figure}

The quasi-particle dispersion $E_{{\bf k} \pm}$ is plotted in
Fig.~\ref{spectrum}. The spectrum is sensitive to the polar angle $\theta$ of the momentum vector ${\bf k}$. For ${\bf k}$ along the $z$-axis (i.e., $\theta=0$ and $k_\perp =0$), $E_{{\bf k}-}$ may become zero at certain values of $k$ when $h$ is sufficiently large, signaling a gapless dispersion. The points at which $E_{{\bf k}-}=0$ are called Fermi points. The value of the Zeeman field at which new Fermi points appear represents a quantum critical point for topological phase transition \cite{chuanwei1,melo1}. For $\mu >0$, the system may support 0, 2 or 4 Fermi points along the $k_z$-axis as $h$ is increased. For $\mu <0$ which is the case illustrated in Fig.~\ref{spectrum}, there can be either 0 or 2 Fermi points \cite{yi}. For $\lambda k_{F}/E_{F}=2$, the critical Zeeman field is $h_c \approx 1.5 E_F$. For $h< h_c$, the system is a topologically trivial gapped superfluid; for $h>h_c$, the system possesses two Fermi points and represents a topologically nontrivial gapless superfluid. Note that the quasi-particle dispersion has been measured in recent experiments on ultracold Fermi gases \cite{jin,jin1}.

\begin{figure}[h]
\includegraphics[width=8.5cm]{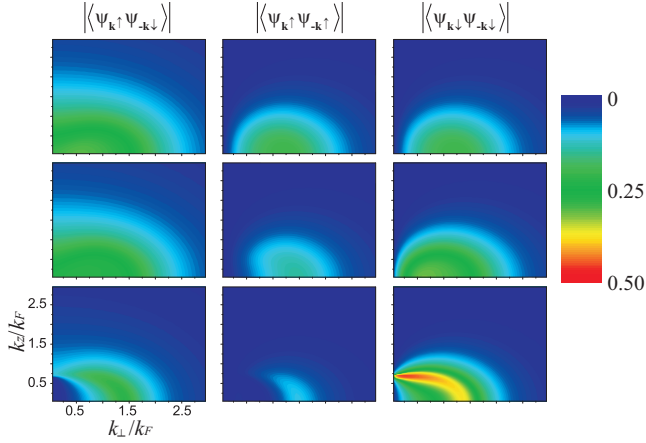}\caption{(color online). Pairing fields at unitarity (i.e., $1/a_s=0$) for $h=0$ (top row), $h=1E_F$ (middle row), $h=2E_{F}$ (bottom row) and $\protect\lambda k_{F}/E_{F}=2$. In each row, from left to right, we display $|\langle
\protect\psi _{{\bf k}\uparrow }\protect\psi _{-{\bf k}\downarrow }\rangle |$,
$|\langle \protect\psi _{{\bf k}\uparrow }\protect\psi %
_{-{\bf k}\uparrow }\rangle| $ and $|
\langle \protect\psi _{{\bf k}\downarrow }\protect\psi _{-{\bf k}\downarrow }\rangle |$ (all in units of $E_F$), respectively.}%
\label{pairing}%
\end{figure}

Even though in our model the interaction has a contact $s$-wave form, due to the presence of the SO coupling, pairing can occur in both singlet and triplet channels \cite{gorkov}.
The singlet pairing field between unlike spins can be calculated as:
\begin{widetext}
\begin{eqnarray}
\langle \psi _{{\bf k}\uparrow }\psi _{-{\bf k}\downarrow }\rangle &=&-\Delta_0
\left\{ \sum_{\alpha=\pm}\frac{2f(E_{{\bf k}\alpha})-1}{4E_{{\bf k}\alpha}}[1+\alpha\frac{h^{2}}{\sqrt{%
(h^{2}+\lambda ^{2}k_{\perp }^{2})\xi _{\bf k}^{2}+h^{2}\Delta_0 ^{2}}}] \right\} \,,
\end{eqnarray}%
while the triplet pairing fields between like spins are given by:
\begin{eqnarray}
\langle \psi _{{\bf k}\uparrow }\psi _{-{\bf k}\uparrow }\rangle &=& \Delta_0 \frac{\lambda
(k_{y}+ik_{x})(\xi _{\bf k}-h)}{\sqrt{(h^{2}+\lambda ^{2}k_{\perp }^{2})\xi
_{\bf k}^{2}+h^{2}\Delta_0 ^{2}}}\left[\frac{2f(E_{{\bf k}+})-1}{4E_{{{\bf k}+}}}-\frac{%
2f(E_{{\bf k}-})-1}{4E_{{\bf k}-}}\right] \,,\\
\langle \psi _{{\bf k}\downarrow }\psi _{-{\bf k}\downarrow }\rangle & =&-\Delta_0 \frac{%
\lambda (k_{y}-ik_{x})(\xi _{\bf k}+h)}{\sqrt{(h^{2}+\lambda ^{2}k_{\perp
}^{2})\xi _{\bf k}^{2}+h^{2}\Delta_0 ^{2}}}\left[\frac{2f(E_{{\bf k}+})-1}{4E_{{\bf k}+}%
}-\frac{2f(E_{{\bf k}-})-1}{4E_{{\bf k}-}}\right]\,.
\end{eqnarray}
\end{widetext}
We plot in Fig.~\ref{pairing} the absolute value of the various pairing fields. One can see that the effect of the Zeeman field is to reduce the singlet pairing while enhancing the triplet pairing.

From the single-particle Green function, one can immediately obtain the density of states which is an important quantity characterizing the nature of the quantum state. To this end, we invert Eq.~(\ref{gf0}) to get
\begin{equation}
{\cal G}_0\left( {\bf k},i\omega _m\right) =\left[
\begin{array}{cc}
\hat{g}({\bf k},i\omega _m) & \hat{f}({\bf k},i\omega _m) \\
\left[ \hat{f}(+{\bf k},-i\omega _m)\right] ^{+} & -\left[ \hat{g}(-{\bf k}%
,-i\omega _m)\right] ^T
\end{array}
\right] ,
\label{g0}
\end{equation}
where $\hat{g}$ and $\hat{f}$ are both $2 \times 2$ matrices whose expressions can be found Appendix~\ref{spg}.
The density of states are related to the $\hat{g}$ matrix as
\begin{eqnarray}
\rho _{\uparrow }(\omega) &=&-\frac{1}{\pi } {\rm Im%
} \left[\frac{1}{V} \underset{\bf k}{\sum } g_{11}(k,\omega+i0^{+}) \right] \,, \\
\rho _{\downarrow }(\omega) &=&-\frac{1}{\pi }  {\rm %
Im} \left[\frac{1}{V} \underset{\bf k}{\sum } g_{22}(k,\omega+i0^{+}) \right]\,.
\end{eqnarray}
where the expressions of $g_{11}$ and $g_{22}$ can be found in Eqs.~(\ref{g11}) and (\ref{g22}) in the Appendix, respectively.

\begin{figure}[h]
\includegraphics[width=8 cm]{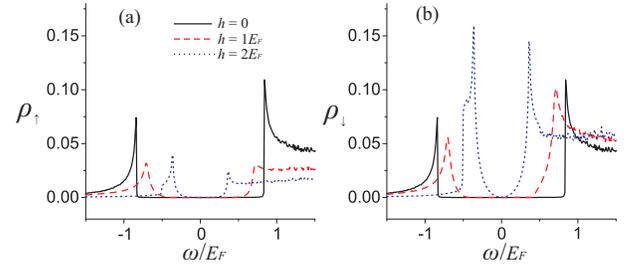}\caption{(color online). Density of states $\rho_\uparrow$ (a) and $\rho_\downarrow$ (b) at unitarity and $\lambda k_F/E_F=2$.}%
\label{dos}%
\end{figure}

We show in Fig.~\ref{dos} the density of states at different Zeeman
fields. At $h=0$, $\rho_\uparrow$ and $\rho_\downarrow$ are
identical, both exhibiting a large gapped region. As $h$ increases,
the gapped region shrinks. Furthermore, the density of states for
the majority component (which is the spin down component for our
choice of positive $h$) becomes more V-shaped near $\omega=0$. A
V-shaped density of states is characteristic of many unconventional
superconductors featuring non-$s$-wave pairing. Here it is due to
the triplet pairing resulting from the SO coupling and enhanced by
the Zeeman field. The density of states may be measured in
experiment using the scheme proposed in Ref.~\cite{lei}.

\begin{figure}[tbh]
\includegraphics[width=8cm]{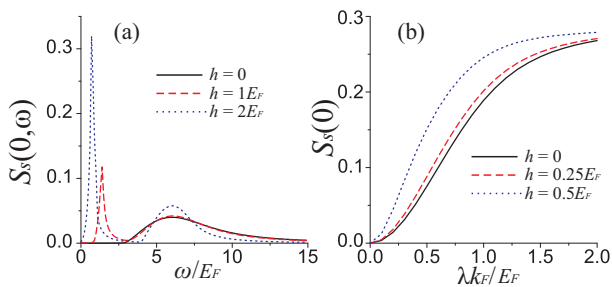}
\caption{(color online). (a) Zero temperature dynamic spin structure factor $S_S(0,\omega)$ at unitarity and $\lambda k_F/E_F=2$. (b) Static spin structure factor $S_S(0)$ as functions of the SO coupling strength.}%
\label{structure factor}%
\end{figure}

Finally,
let us consider spin structure factor \cite{factor1,factor} which is related to the dynamic spin susceptibility $\chi _{S}\left( \mathbf{q}%
,i\nu _{n}\right) $, which is the Fourier transformation of the spin-spin
correlation function
\begin{equation}
\chi _{S}\left( \mathbf{x},\tau \right) =-\left\langle {\rm T}_{\tau }\delta
n_{S}\left( \mathbf{x},\tau \right) \delta n_{S}\left( \mathbf{0},0\right)
\right\rangle \,.
\end{equation}
where the spin density is given by $\delta n_{S}=n_{\uparrow }-n_{\downarrow }$.
Using the Nambu spinor notation, the spin density can be written as,
\begin{equation}
\delta n_{S}=\frac{1}{2}\Phi ^{+}\left[
\begin{array}{cccc}
1 & 0 & 0 & 0 \\
0 & -1 & 0 & 0 \\
0 & 0 & -1 & 0 \\
0 & 0 & 0 & 1%
\end{array}%
\right] \Phi =\frac{1}{2}\Phi ^{+}\left[ \hat{\tau}_{z}\otimes \hat{\sigma}%
_{z}\right] \Phi .
\end{equation}%
where $\hat{\tau}_i$ are the Pauli matrices describing the Nambu spinor degrees of freedom.
Within the static Bogoliubov approximation \cite{note}, we have,
\begin{widetext}
\begin{equation}
\chi _{S}\left( \mathbf{q},i\nu _{n}\right) =\frac{1}{4}k_{B}T\sum_{\mathbf{k%
},i\omega _{m}}\mathrm{Tr}\left\{ \hat{\tau}_{z}\otimes \hat{\sigma}_{z}%
\mathcal{G}_{0}\left( \mathbf{k},i\omega _{m}\right) \hat{\tau}_{z}\otimes
\hat{\sigma}_{z}\mathcal{G}_{0}\left( \mathbf{k+q},i\omega _{m}+i\nu
_{n}\right) \right\} .
\end{equation}%
The zero-momentum dynamic spin structure factor is given by
\begin{equation}
S_{S}\left( \mathbf{q=0},\omega \right) =-\frac{1}{1-e^{-\omega /k_{B}T}}%
\frac{1}{\pi } {\rm Im} \left[ \chi _{S}\left( \mathbf{q=0},i\nu _{n}\rightarrow
\omega +i0^{+}\right) \right] \,,
\end{equation}%
\end{widetext}
whose analytic expression is too complicated to show. However it can be explicitly shown that the dynamic spin structure factor vanishes when $\lambda=0$, i.e., in the absence of the SO coupling. Fig.~\ref{structure factor}(a) shows dynamic spin structure factor at
different Zeeman fields. At $h=0$, $S_S(0,\omega)$ exhibits a broad peak. For finite $h$, an additional narrower peak appears at smaller energy. The corresponding static spin structure factor is given by \[ S_S(0) = \int d\omega \,S_S(0,\omega) \,,\] which is plotted in Fig.~\ref{structure factor}(b) as a function of the SO coupling strength. The spin structure factor may be directly measured in experiment using the Bragg spectroscopic method \cite{bragg}.

\section{Outlook and Conclusion}
\label{oc} In this paper, we provide a detailed theoretical
description of a two-component Fermi gas with Rashba SO coupling,
under the framework of the functional path integral formalism. Many
important and experimentally relevant physical quantities in both
two- and many-body situations --- such as the effective mass of the
two-body bound state, the chemical potential, gap parameter, pairing
correlation, quasi-particle dispersion, density of states, spin
structure factor, etc. --- are calculated in a rather
straightforward way using this method. The SO coupling in general
favors the formation of Cooper pair and bound state. In particular,
we have found that when a Rashba ring exists for the single-particle
ground state, a two-body bound state exists regardless of the sign
of the $s$-wave scattering length.

For the sake of simplicity, we have only presented the zero-temperature results in this paper. The formalism and the relevant equations we have derived are nevertheless valid for finite temperatures.
In addition, we have focused on a uniform system. To take the trap into account, one may adopt the local density approximation \cite{hu,yi}, under which the trapped Fermi gas was
treated as the sum of many small cells with a local chemical potential $\mu
({\bf r})=\mu -V_{\rm trap}({\bf r})$. In our earlier work \cite{hu}, we have shown that the important qualitative features of the system, like the mixed-spin pairing, anisotropy, enhanced pairing field, etc., are not affected by the presence of the trap. A  self-consistent way to include the effects of the trapping potential is to use the Bogoliubov-de Gennes (BdG) formalism. For large number of atoms as used in most experiments, we expect that the local density approximation should be quite accurate. However, for a system with relatively small number of atoms and tight trapping confinement, or in the case of vortex state where the order parameter varies significantly in a short length scale, the BdG approach will be more appropriate.

One crucial feature of the system arising from the SO coupling is
that the superfluid transition temperature is greatly enhanced
\cite{zhai}. The mean-field calculation which we have focused on in
this work is not expected to provide an accurate estimate of the
transition temperature, particularly for system with strong
interaction strength. As we have outlined in Sec.~\ref{formalism} of
this paper, the Gaussian fluctuations on top of the mean-field level
can be accounted for using the functional path integral formalism.
This will lead to a much more accurate calculation of the transition
temperature. We plan to address this issue in a future work.

\begin{acknowledgments}
HP is supported by the NSF and the Welch Foundation (Grant No. C-1669). HH and XJL was supported by the ARC Discovery
Projects No. DP0984522 and No. DP0984637. We would like to thank Chuanwei Zhang, Hui Zhai and Shizhong Zhang for useful discussions.
\end{acknowledgments}

\appendix
\begin{widetext}
\section{Expressions of Single-Particle Green Function}
\label{spg}
The single-particle Green function is given in Eq.~(\ref{g0}) where the explicit expressions of the elements are
\begin{eqnarray*}
\hat{g}({\bf k},i\omega _m) &=&\left[ i\omega _m+\xi _{{\bf k}}-h\hat{\sigma}%
_z+\lambda (k_y\hat{\sigma}_x-k_x\hat{\sigma}_y)\right] {\cal D}_{+}/M, \\
\hat{f}({\bf k},i\omega _m) &=&\left[ -i\Delta _0\hat{\sigma}_y\right] {\cal %
D}_{-}/M.
\end{eqnarray*}
with
\begin{equation}
M=\left[ \left( i\omega _m-h\right) ^2-\xi _{{\bf k}}^2-\Delta _0^2-\lambda
^2k_{\perp }^2\right] \left[ \left( i\omega _m+h\right) ^2-\xi _{{\bf k}%
}^2-\Delta _0^2-\lambda ^2k_{\perp }^2\right] -4\lambda ^2k_{\perp }^2\left(
\xi _{{\bf k}}^2-h^2\right) ,
\end{equation}
and
\begin{eqnarray}
{\cal D}_{+} &=&\left[
\begin{array}{cc}
\left( i\omega _m+h\right) ^2-\xi _{{\bf k}}^2-\Delta _0^2-\lambda
^2k_{\perp }^2 & \lambda \left( k_y+ik_x\right) \left[ 2\left( \xi _{{\bf k}%
}+h\right) \right] \\
\lambda \left( k_y-ik_x\right) \left[ 2\left( \xi _{{\bf k}}-h\right) \right]
& \left( i\omega _m-h\right) ^2-\xi _{{\bf k}}^2-\Delta _0^2-\lambda
^2k_{\perp }^2
\end{array}
\right] , \\
{\cal D}_{-} &=&\left[
\begin{array}{cc}
\left( i\omega _m-h\right) ^2-\xi _{{\bf k}}^2-\Delta _0^2-\lambda
^2k_{\perp }^2 & -\lambda \left( k_y-ik_x\right) \left[ 2\left( \xi _{{\bf k}%
}+h\right) \right] \\
-\lambda \left( k_y+ik_x\right) \left[ 2\left( \xi _{{\bf k}}-h\right)
\right] & \left( i\omega _m+h\right) ^2-\xi _{{\bf k}}^2-\Delta _0^2-\lambda
^2k_{\perp }^2
\end{array}
\right] .
\end{eqnarray}
In greater detail, we find that,
\begin{equation}
\hat{g}({\bf k},i\omega _m)=\left[
\begin{array}{cc}
\hat{g}_{11}\left( {\bf k},i\omega _m\right) & \hat{g}_{12}\left( {\bf k}%
,i\omega _m\right) \\
\hat{g}_{12}^{*}\left( {\bf k},-i\omega _m\right) & \hat{g}_{22}\left( {\bf k%
},i\omega _m\right)
\end{array}
\right] ,
\end{equation}
where,
\begin{eqnarray}
M\hat{g}_{11}\left( {\bf k},i\omega _m\right) &=&\left[ i\omega _m+\xi _{{\bf k}}-h\right]
\left[ \left( i\omega _m+h\right) ^2-\xi _{{\bf k}}^2-\Delta _0^2-\lambda
^2k_{\perp }^2\right] +2\lambda ^2k_{\perp }^2\left( \xi _{{\bf k}}-h\right)
,\label{g11} \\
M\hat{g}_{12}\left( {\bf k},i\omega _m\right) &=&\lambda \left( k_y+ik_x\right) \left[ \left(
i\omega _m+\xi _{{\bf k}}\right) ^2-h^2-\Delta _0^2-\lambda ^2k_{\perp
}^2\right] , \\
M\hat{g}_{22}\left( {\bf k},i\omega _m\right) &=&\left[ i\omega _m+\xi _{{\bf k}}+h\right]
\left[ \left( i\omega _m-h\right) ^2-\xi _{{\bf k}}^2-\Delta _0^2-\lambda
^2k_{\perp }^2\right] +2\lambda ^2k_{\perp }^2\left( \xi _{{\bf k}}+h\right)
, \label{g22}
\end{eqnarray}
and
\begin{equation}
\hat{f}({\bf k},i\omega _m)=\frac{\Delta _0}M\left[
\begin{array}{cc}
\lambda \left( k_y+ik_x\right) \left[ 2\left( \xi _{{\bf k}}-h\right) \right]
& -\left[ \left( i\omega _m+h\right) ^2-\xi _{{\bf k}}^2-\Delta _0^2-\lambda
^2k_{\perp }^2\right] \\
+\left[ \left( i\omega _m-h\right) ^2-\xi _{{\bf k}}^2-\Delta _0^2-\lambda
^2k_{\perp }^2\right] & -\lambda \left( k_y-ik_x\right) \left[ 2\left( \xi _{%
{\bf k}}+h\right) \right]
\end{array}
\right] .
\end{equation}
\end{widetext}


\begin{thebibliography}{99}

\bibitem{lin1}Y.-J. Lin, R. L. Compton, A. R. Perry, W. D. Phillips, J. V. Porto, and I. B. Spielman, Phys. Rev. Lett. {\bf 102}, 130401 (2009).

\bibitem{lin2}Y.-J. Lin, R. L. Compton, K. Jim\'{e}nez-Garc\'{i}a, J. V. Porto, and I. B. Spielman, Nature (London) {\bf 462}, 628 (2009).

\bibitem{lin3}Y-J. Lin, R. L. Compton, K. Jim\'{e}nez-Garc\'{i}a, W. D. Phillips, J. V. Porto, and I. B. Spielman, Nature Phys. {\bf 7}, 531 (2011).

\bibitem{lin4}Y.-J. Lin, K. Jim\'{e}nez-Garc\'{i}a, and I. B. Spielman, Nature (London) \textbf{471}, 83 (2011).

\bibitem{zhang}Z. Fu, P. Wang, S. Chai, L. Huang, and J. Zhang, e-print arXiv:1106.0199.

\bibitem{dalibard}For a brief review, see, for example, Jean Dalibard, Fabrice Gerbier, Gediminas Juzeli\={u}-nas, Patrik \"{O}hberg, eprint arXiv:1008.5378.

\bibitem{shenoy}J. P. Vyasanakere, and V. B. Shenoy, Phys. Rev. B {\bf 83},
094515 (2011).

\bibitem{shenoy1}J. P. Vyasanakere, S. Zhang, and V. B. Shenoy,Phys. Rev. B {\bf 84}, 014512 (2011).

\bibitem{hu}H. Hu, L. Jiang, X.-J. Liu, and H. Pu, e-print arXiv:1105.2488 (to appear in Phys. Rev. Lett.)

\bibitem{zhai}Z.-Q. Yu, and H. Zhai, e-print arXiv:1105.2250.

\bibitem{chuanwei} C. Zhang, S. Tewari, R.M. Lutchyn, and S. Das Sarma, Phys. Rev. Lett. {\bf 101}, 160401 (2008).

\bibitem{chuanwei1}M. Gong, S. Tewari, and C. Zhang, e-print arXiv:1105.1796 (to appear in Phys. Rev. Lett.)

\bibitem{yi}W. Yi, and G.-C. Guo, e-print arXiv:1106.5667.

\bibitem{sau}J.D. Sau, R. Sensarma, S. Powell, I. B. Spielman, and S. Das Sarma, Phys. Rev. B {\bf 83}, 140510(R) (2011)

\bibitem{iskin}M. Iskin, and A. L. Suba\c{as}, Phys. Rev. Lett. {\bf 107}, 050402 (2011).

\bibitem{iskin1}M. Iskin, and A. L. Suba\c{as}, e-print arXiv:1108.4263.

\bibitem{baym}T. Ozawa, and G. Baym, e-print arXiv:1107.3162.

\bibitem{melo}L. Han, and C. A. R. S\'{a} de Melo, e-print arXiv:1106.3613.
\bibitem{melo1}K. Seo, L. Han, and C. A. R. S\'{a} de Melo, e-print arXiv:1108.4068.

\bibitem{two1}L. Dell'Anna, G. Mazzarella, and L. Salasnich, Phys. Rev. A {\bf 84}, 033633 (2011).

\bibitem{two2}G. Chen, M. Gong, and C. Zhang, e-print arXiv:1107.2627.

\bibitem{two3}B. Huang, and S. Wan, e-print arXiv:1109.3970.

\bibitem{two4}L. He, and X.-G. Huang, e-print arXiv:1109.5577.

\bibitem{gorkov}L.P. Gor'kov, and E.I. Rashba, Phys. Rev. Lett. {\bf 87}, 037004 (2001).

\bibitem{tewari}S. Tewari, T.D. Stanescu, J.D. Sau, and S. Das Sarma, New J. Phys. {\bf 13}, 065004 (2011).

\bibitem{sato}M. Sato, Y. Takahashi, and S. Fujimoto, Phys. Rev. Lett. {\bf 103}, 020401 (2009); Phys. Rev. B {\bf 82}, 134521 (2010).

\bibitem{sademelo} C. A. R. S\'{a} de Melo, M. Randeria, and J. R.
Engelbrecht, Phys. Rev. Lett. \textbf{71}, 3202 (1993); M. Randeria,
in \emph{Bose-Einstein Condensation}, edited by A. Griffin, D. W.
Snoke, and S. Stringari, (Cambridge University Press, Cambridge,
England, 1995), p. 355-392.

\bibitem{hu2} H. Hu, X.-J. Liu, and P. Drummond, Europhys. Lett. \textbf{74}, 574
(2006); R. B. Diener, R. Sensarma, and M. Randeria, Phys. Rev. A
\textbf{77}, 023626 (2008).

\bibitem{nsr} P. Nozières and S. Schmitt-Rink, J. Low Temp. Phys.
\textbf{59}, 195 (1985).

\bibitem{Liu} X.-J. Liu and H. Hu, Europhys. Lett. \textbf{75}, 364
(2006).

\bibitem{jin}J. T. Stewart, J. P. Gaebler, and D. S. Jin, Nature (London) {\bf 454}, 744 (2008).
\bibitem{jin1}J. P. Gaebler, J. T. Stewart, T. E. Drake, D. S. Jin, A. Perali, P. Pieri, and G. C. Strinati, Nature Phys. {\bf 6}, 569 (2010).

\bibitem{lei}L. Jiang, L. O. Baksmaty, H. Hu, Y. Chen, and H. Pu, Phys. Rev. A {\bf 83}, 061604(R) (2011).

\bibitem{factor1}A. Minguzzi, G. Ferrari, and Y. Castin, Eur. Phys. J. D {\bf 17} 49, 2001.

\bibitem{factor}R. Combescot, S. Giorgini, and S. Stringari, Europhys. Lett. {\bf 75} 695 (2006).

\bibitem{note}One may obtain a more accurate evaluation of the structure factor using the random phase approximation \cite{factor1}. This will be studied in the future.

\bibitem{bragg}G. Veeravalli, E. Kuhnle, P. Dyke, and C. J. Vale, Phys. Rev. Lett. {\bf 101}, 250403 (2008); E. D. Kuhnle, H. Hu, X.-J. Liu, P. Dyke, M. Mark, P. D. Drummond, P. Hannaford, and C. J. Vale, Phys. Rev. Lett. {\bf 105}, 070402 (2010); E. D. Kuhnle, S. Hoinka, P. Dyke, H. Hu, P. Hannaford, and C. J. Vale, Phys. Rev. Lett. {\bf 106}, 170402 (2011).

\end{thebibliography}
\end{document}